\documentclass[10pt]{article} 
\usepackage{amssymb}
\usepackage{latexsym} 
\usepackage[dvips]{epsfig}

\newtheorem{theorem}{Theorem}

\def\ethbar{\overline{\eth}}

\def\pib{\overline{\pi}} 
\def\rhob{\overline{\rho}}
\def\mub{\overline{\mu}} 
\def\alphab{\overline{\alpha}}
\def\mbar{\overline{m}} 
\def\xib{\overline{\xi}}
\def\deltab{\overline{\delta}}

\def\O{\mathcal{O}}
\def\caln{\mathcal{N}} 
\def\calz{\mathcal{Z}} 
\def\cals{\mathcal{S}}

\def\gtilde{\widetilde{g}}

\def\phit{\widetilde{\phi}} 
\def\calnt{\widetilde{\caln}}
\def\phii{\stackrel{(1)}{\phi}} 
\def\phio{\stackrel{(0)}{\phi}}
\def\phiit{\widetilde{\phii}} 
\def\phiot{\widetilde{\phio}}

\font\SYM=msbm10 
\newcommand{\Real}{{\SYM R}}

\font\tenscr=rsfs10 scaled1100
\font\sevenscr=rsfs7 
\font\fivescr=rsfs5 
\skewchar\tenscr='177 
\skewchar\sevenscr='177
\skewchar\fivescr='177 
\newfam\scrfam 
\textfont\scrfam=\tenscr
\scriptfont\scrfam=\sevenscr 
\scriptscriptfont\scrfam=\fivescr

\def\scri{{\fam\scrfam I}} 
\def\scrm{{\fam\scrfam M}}
\def\scru{{\fam\scrfam U}}

\def\scrmtilde{\widetilde{\scrm}}

\begin{document}

\title{On the existence and convergence of polyhomogeneous expansions of zero-rest-mass fields.}  
\author{Juan Antonio Valiente Kroon \thanks{E-mail address:
{\tt j.a.valiente@qmw.ac.uk}} \\
School of Mathematical Sciences,\\ Queen Mary \& Westfield College,\\
Mile End Road, London E1 4NS,\\ United Kingdom.}

\maketitle

\begin{abstract}
The convergence of polyhomogeneous expansions of zero-rest-mass fields
in asymptotically flat spacetimes  is discussed. An existence proof
for the asymptotic characteristic initial value problem for a
zero-rest-mass field with polyhomogeneous initial data is given. It is
shown how this non-regular problem can be properly recast as a set of
regular initial value problems for some auxiliary fields. The standard
techniques of symmetric hyperbolic systems can be applied to these new
auxiliary problems, thus yielding a positive answer to the
question of existence in the original problem.

\end{abstract}

\section{Introduction}
The analysis of the asymptotic behaviour of asymptotically flat
spacetimes and the diverse fields propagating on it has been carried out by
assuming that both the gravitational field and the test fields
(e.g. zero-rest-mass fields) are smooth at the boundary of the
\emph{unphysical} (i.e. conformally rescaled) spacetime manifold (null
infinity, $\scri$) \cite{Pen63,Pen65a,PenRin86}. This smoothness at
null infinity implies that the fields in the physical spacetime
manifold have the peeling behaviour. However, this assumption may be
too strong for a number of physical applications (see for example
\cite{ChrMacSin95} and references within). In particular, the case for 
the so-called \emph{polyhomogeneous spacetimes
and zero-rest-mass fields} has been rised 
\cite{AndChr94,ChrMacSin95,Val98,Val99a,Val99b,Val00a}. A generic
peeling (smooth) field possesses an analytic expansion around null infinity
(given by $\Omega=0$). By contrast, the expansions of polyhomogeneous
fields around $\scri$ are Laurent-like (so that the field in the
physical spacetime is non-peeling), and also contain powers of $\ln$.

So far, the study of polyhomogeneous fields has been carried out
assuming that the diverse fields admit expansions of the required
form. However, the expansions are formal in the sense that the
convergence (and hence the existence and uniqueness) of the series has
not been discussed. This point, so many times overlooked, is of
fundamental importance. It could be well the case that none of the
formal expansions converge!

The objective of this article is to provide a first step towards the
understanding of the convergence of polyhomogeneous fields. Here, I
analyse the possible convergence of a polyhomogeneous spin-1
zero-rest-mass field (Maxwell field) propagating on an asymptotically
simple spacetime. The spin-1 zero-rest-mass field has been chosen in
order to ease the discussion and the calculations, but the treatment
here discussed can be directly extended to the case of say, a spin-2
field. It is also expected that similar techniques can be used to
analyse the analogous question for the gravitational field. However,
in that case one should expect a number of complications. Furthermore,
again for concreteness, only the simplest case of spin-1
zero-rest-mass fields will be discussed: the so-called minimally
polyhomogeneous. This case contains all the difficulties one could
expect from the generic polyhomogeneous situation, but the algebra is
much simpler!

The question of the convergence of the polyhomogeneous expansions will
be tackled by posing an \emph{asymptotic characteristic initial value
problem} for the minimally polyhomogeneous spin-1 zero-rest-mass
field, and proving that such a problem has a (unique) solution in a
given neighbourhood of the spacetime which is polyhomogeneous
(i.e. the solution has convergent polyhomogeneous expansion in the
given neighbourhood).

\section{Preliminaries, notation, conventions}
Let $(\scrmtilde,\gtilde_{ab})$ be an asymptotically simple
spacetime. It will be assumed that null infinity is smooth
(non-polyhomogeneous). Let $(\scrm,g_{ab})$ be the unphysical
spacetime obtained by means of the conformal rescaling
$g_{ab}=\Omega^2\gtilde_{ab}$. The spacetime will be taken to be
completely known. In particular this means that the asymptotic
expansions of the spin coefficients, and the tetrad functions are
known. Let $\caln_0$ be the initial (future oriented) null
hypersurface, and let $\calz_0$ be its intersection with future null
infinity, $\calz_0=\caln_0\cap\scri^+$. Following Penrose \& Rindler's
notation, $\approx$ will denote equality at null infinity.

To describe the spacetime and the different fields involved, I will
make use of the coordinates and null tetrad described by Friedrich \&
Stewart. In their original version, these coordinates and null tetrad
were used to describe the spacetime between two intersecting null
hypersurfaces \cite{SteFri82,Ste91}. Here they are adapted to describe
the region of spacetime in the causal past of the intersection of
future null infinity ($\scri^+$), and a future oriented null
hypersurface. The construction is of interest in itself, and its
highlights are given for the sake of reference.

The physical spacetime $(\scrmtilde,\gtilde_{ab})$ is asymptotically
flat, thus $\calz_0$ has the topology of the 2-sphere, $\cals^2$. Let
$p$ be an arbitrary point of $\calz_0$, and let $\gamma$ and $\gamma'$
be the null generators of $\caln_0$ and $\scri^+$ through $p$
respectively. A null tetrad $\{l^a,n^a,m^a,\mbar^a\}$ at $p$ can be
chosen such that $m^a$ and $\mbar^a$ span $\mbox{T}_{p}(\calz_0)$, and
$l^a$ and $n^a$ are tangent to $\gamma$ and $\gamma'$
respectively. One can also choose an arbitrary coordinate system,
$(x^\alpha)$, at $\calz_0$. Let $u$ (retarded time) be a parameter
along the null geodesic generators $\gamma'$ of $\scri^+$, such that
$u=0$ at $p$. Note that in principle $u$ is not an affine
parameter. Repeating this construction for different points,$p$, one
obtains a scalar field  that vanishes precisely at $\calz_0$. The
coordinates $(x^\alpha)$ can now be extended to the remainder of
$\scri^+$ by requiring the coordinates to be constant along the
generators of $\scri^+$. In this way we have constructed a coordinate
chart $(u,x^\alpha)$ for $\scri^+$ valid at least around
$\calz_0$. One can undertake an analogous construction for the
generators $\gamma$ of $\caln_0$, obtaining a parameter $v$ (advanced
time) such that at $p\in\calz_0$, $v=0$. Again, the parameter $v$ is
not affine in principle. Thus, one ends up with a coordinate chart
$(v,x^\alpha)$ for $\caln_0$.

Now, on $\scri^+$ there is a 1-parameter family of spacelike
hypersurfaces, $\cals_u$ (cuts) on which $u$ is constant. For each
$\cals_u$ there is a unique null hypersurface $\caln_u$ containing
$\cals_0$ ($\cals_u\subset\caln_u$) extending into the interior of the
unphysical spacetime. This yields a foliation of the spacetime by null
hypersurfaces at least close to $\scri^+$ ---note that the foliation
could give rise to caustics and cusps. Similarly, one can obtain
another 1-parameter family of null hypersurfaces $\caln'_v$ for which
$v$ is constant and that intersect $\caln_0$. In particular one has
that $\caln_0'=\scri^+$. The coordinates $(x^\alpha)$ can now also be
extended to the interior of the spacetime by requiring them to be
constant along the null generators of $\caln_u$. Note that
$(x^\alpha)$ are not constant along the generators of the $\caln_v'$
foliation (except for $\scri^+$) for the null vectors $l^a$ and $n^a$
which will be taken as tangent to the null generators of $\caln_u$ and
$\caln_v'$ respectively do not in general commute.

To extend the null tetrad defined in $\calz_0$, one begins by defining
$l_a$ on $\scrm$ by $l_a=\nabla_au$. Therefore $l^a$ is tangent to the
null generators of $\caln_0$, and is parallel to the $l^a$ already
defined at $\calz_0$. A boost on the null tetrad defined at $\calz_0$
can be used to make the two vectors equal. Thus, there exists a scalar
function $Q$ such that $l^a\partial_a=Q\partial_v$. It can be shown
that $Q=\O(1)$. Next, the $n_a$ is defined on $\scrm$ by
$n_a=Q^{-1}\partial_v$, so that $n^a$ is tangent to the null
generators of $N_v'$ and the normalisation condition $l^an_a=1$ is
satisfied. Together with $n^an_a=0$ this implies,
$n^a\partial_a=\partial_u+c^\alpha\partial_\alpha$. Due to the
constancy of the $(x^\alpha)$ coordinates along the generators of
$\scri^+$, one has that $C^\alpha\approx0$.

Following the standard NP notation, let $D=l^a\nabla_a$ and
$\Delta=n^a\nabla_a$. The vector $m^a$ defined so far only on
$\calz_0$ is propagated onto the remaining of $\scri^+$ via the
equation $\Delta m^a=-\tau n^a$. Finally $m^a$ is propagated from
$\scri^+$ to the interior of $\scrm$ via $Dm^a=-\pib l^a$. This together
with the nullity and orthogonality requirements imply that
$m^a\partial_a=\xi^\alpha\partial_\alpha$.

Hence, the aforediscussed null tetrad has the form:
\begin{eqnarray}
&&l^a=Q\delta^a_v, \\ &&n^a=\delta^a_u + C^\alpha\delta_\alpha^a, \\
&&m^a=\xi^\alpha\delta_\alpha^a, \\
&&\mbar^a=\xib^\alpha\delta_\alpha^a,
\end{eqnarray}
where $C^\alpha\approx0$, and $Q$, $\xi^\alpha$ are of order
$\O(1)$. From this tetrad construction one also deduces that
$\kappa=\nu=0$, $\rho=\rhob$, $\mu=\mub$, $\epsilon=0$,
$\tau=\alphab+\beta$ in $\scrm$. In particular one also has that all
the spin coefficients with the exception of $\sigma$ vanish at
$\scri^+$: that is, they are of order $\O(\Omega)$. On the other hand,
$\sigma$ is $\O(1)$.

Finally, notice that the  aforediscussed construction fixes also the
associated spin basis $\{o^A,\iota^A\}$.

\subsection{Polyhomogeneous zero-rest-mass fields}
 A spin-s zero-rest-mass field is a totally symmetric spinor field
 ($\phi_{AB\cdots C}=\phi_{(AB\cdots C)}$), satisfying:
\begin{equation}
\nabla^{AA'}\phi_{AB\cdots C}=0,
\end{equation}
where the indices $AB\cdots C$ number $2s$. In order to ease the
discussion, in this article we will restrict ourselves to spin 1
zero-rest-mass fields (Maxwell field). The extension of the results to
spin 2 fields (linear gravity) is direct. The components of the spin 1
zero-rest-mass field in the aforementioned spin basis are:
$\phi_0=\phi_{AB}o^Ao^B$, $\phi_1=\phi_{AB}o^A\iota^B$,
$\phi_2=\phi_{AB}\iota^A\iota^B$.

Let us consider zero-rest-mass fields belonging to
$C^\infty_{loc}(\scrm)$, i.e. fields that are infinite differentiable
(smooth) in the interior of the unphysical spacetime
($\stackrel{\circ}{\scrm}=\scrmtilde$) but need not to be so at the
boundary ($\scri$) of the spacetime; this is the meaning of the
$loc$ label. In particular our attention will be centered on the
so-called \emph{polyhomogeneous fields}, which are
$C^\infty_{loc}(\scrm)$ zero-rest-mass fields with asymptotic
expansions in terms of powers of a particular parameter (e.g. the
conformal factor, and affine parameter of outgoing null geodesics, a
luminosity parameter, an advanced time), and powers of the logarithm. 

The most general polyhomogeneous spin-1 zero-rest-mass field is of the form,
\begin{eqnarray}
&&\phit_0=\phi_0^2\Omega^2+\phi_0^3\Omega^3+\cdots,  \label{poly1}\\
&&\phit_1=\phi_1^2\Omega^2+\phi_1^3\Omega^3+\cdots,  \label{poly2}\\
&&\phit_2=\phi_2^1\Omega  +\phi_2^2\Omega^2+\cdots,  \label{poly3}
\end{eqnarray}
where the $\phi_n^i$'s are polynomials in $z=\ln \Omega$
with coefficients in $C^\omega(\mbox{\Real}\times\cals^2)$. It is
customary to require absolute convergence of the polyhomogeneous series
and its derivatives to all orders \cite{ChrMacSin95}.

Note that the field defined by equations (\ref{poly1})-(\ref{poly3})
is clearly non-peeling. A peeling field should behave like
$\phi_n=\O(\Omega^{(3-n)})$. Furthermore, looking at the spin-1
zero-rest-mass field in the unphysical (conformally rescaled)
spacetime one has,
\begin{eqnarray}
&&\phi_0=\phi_0^2\Omega^{-1}+\phi_0^3+\cdots, \\
&&\phi_1=\phi_1^2+\phi_1^3\Omega+\cdots, \\
&&\phi_2=\phi_2^1+\phi_2^2\Omega+\cdots,
\end{eqnarray}
so that the field is not regular (diverges!) at $\scri^+$ (where
$\Omega=0$), first because of the term $\phi_0^2\Omega^{-1}$, and also
because of the presence of logarithmic terms in $\phi_1^2$. 

In order to ease the discussion, the discussion in this article will
be reduced to the case of the so-called minimally polyhomogeneous
fields. An extension to more general polyhomogeneous fields can be
easily done. A minimally polyhomogeneous spin-1 zero-rest-mass field
is of the form (in the physical spacetime):
\begin{eqnarray}
&&\phit_0=\phi_0^{2,0}\Omega^2+\left(\phi_0^{3,1}\ln\Omega+\phi_0^{3,0}\right)\Omega^3+\cdots,
\\
&&\phit_1=\left(\phi_1^{2,1}\ln\Omega+\phi_1^{2,0}\right)\Omega^2+\cdots,
\\ &&\phit_2=\phi_0^{1,0}\Omega+\cdots,
\end{eqnarray}
which in the unphysical spacetime looks like,
\begin{eqnarray}
&&\phi_0=\phi_0^{2,0}\Omega^{-1}+\left(\phi_0^{3,1}\ln
\Omega+\phi_0^{3,0}\right)+\cdots, \\ 
&&\phit_1=\left(\phi_1^{2,1}\ln \Omega+\phi_1^{2,0}\right) +\cdots, \\ 
&&\phit_2=\phi_0^{1,0}+\left(\phi_2^{2,1}\ln \Omega+\phi_2^{2,0}\right)\Omega \cdots.
\end{eqnarray}
Using the ``constraint equations'', i.e. the $D$-Maxwell equations one can deduce several relations connecting the diverse coefficients of the different components of the spin-1 zero-rest-mass field. In particular one has that
\begin{equation}
\phi_1^{2,1}=\ethbar\phi_0^{2,0},
\end{equation}
where $\ethbar$ is the ethbar differential operator (see for example \cite{PenRin84}). This relation will be used later.

So far, the asymptotic expansions have been in $\Omega$. However, it can be seen that,
\begin{equation}
\Omega=a(u,x^\alpha)v+\cdots,
\end{equation}
so that one could alternatively use the retarded time as the expansion
parameter. \emph{This will be done in the remainder of the article}. One can
even redefine $a(u,x^\alpha)$ so that $a(u,x^\alpha)=1$.

A minimally polyhomogeneous field contains at the most first powers of
$\ln$. If one provides $\phit_0$ on an initial null hypersurface
$\calnt_0$, and the coefficients $\phi_1^{2,0}$ at $x^0=0$ and
$\phi_2^{1,0}$ for all times, one can use the equations (\ref{M3}) and
(\ref{M4}) on to obtain the components $\phit_1$ and $\phit_2$ on
$\calnt_0$, and then the evolution equations (\ref{M1}) and (\ref{M2})
to generate the field by ``slices''. This procedure allows us to
calculate formal polyhomogeneous expansions for the zero-rest-mass
field. As mentioned previously in the introduction, the question to be
investigated here regards the convergence of such formal
polyhomogeneous expansions, i.e. we want to know if such fields exist,
at least for a region of the spacetime very close to the initial null
hypersurface $\caln_0$.

\section{Symmetric hyperbolicity and the asymptotic characteristic initial value problem}
The question of the convergence of the polyhomogeneous expansions of the zero-rest-mass fields discussed in the previous section will be addressed by posing a so-called asymptotic characteristic initial value problem, i.e. an initial value problem with initial data given on a future oriented null hypersurface (a characteristic of the zero-rest-mass field equations), and on future null infinity (which, by the way is also a characteristic of the field equations). Once a suitable initial value problem has been posed, an existence result will be proven. This existence result will exhibit the convergence of the polyhomogeneous series. The aforesaid initial value problem is \emph{non-regular} in a sense to be clarified later. Hence the standard techniques used to prove existence results for analytic fields cannot be directly used. Some manipulations will have to be carried. In order to understand the rationale behind the forthcoming discussion, the standard regular problem is briefly reviewed.

\subsection{The initial value problem with regular initial data}

Under the tetrad choice described in the previous sections,  the spin-1 zero rest-mass field equations (aka source-free Maxwell equations) are given by:
\begin{eqnarray}
&&\Delta\phi_0-\delta\phi_1=(2\gamma-\mu)\phi_0-2\tau\phi_1+\sigma\phi_2,
\label{M1}\\
&&\Delta\phi_1-\delta\phi_2=-2\mu\phi_1+(2\beta-\tau)\phi_2, \label{M2}\\
&&D\phi_2-\deltab\phi_1=-\lambda\phi_0+2\pi\phi_1+\rho\phi_2, \label{M3} \\
&&D\phi_1-\deltab\phi_0=(\pi-2\alpha)\phi_0+2\rho\phi_1. \label{M4}
\end{eqnarray}
Recall that the zero-rest-mass fields are totally symmetric, and thus in our case $\phi_{AB}o^A\iota^B=\phi_{AB}\iota^Ao^B$. Therefore the system (\ref{M1})-(\ref{M4}) is overdetermined: 4 equations for 3 complex components. This difficulty is overcome by discarding the last equation (\ref{M4}). Then, one has to verify that this equation is implied by the other ones if it held initially. This can be done by defining,
\begin{equation}
\Phi=D\phi_1-\deltab\phi_0-(\pi-2\alpha)\phi_0-2\rho\phi_1.
\end{equation}
Using the field equations, the commutators and the Ricci identities (NP field equations), it is not difficult to show that,
\begin{equation}
\partial_u\Phi=0.
\end{equation}
Thus, if equation (\ref{M4}) held initially, it will also held for later retarded times.

The principal part (symbol) of the reduced system (\ref{M1})-(\ref{M3}) is contained in the terms
\begin{eqnarray}
&&\Delta\phi_0-\delta_1, \\ &&\Delta\phi_1-\delta_2, \\
&&D\phi_2-\deltab\phi_1,
\end{eqnarray}
and can be written concisely in a matricial way as $A^a\phi_{b,a}$ where
$\phi=(\phi_0,\phi_1,\phi_2)$, and
\begin{equation}
A^u=\left(
\begin{array}{ccc}
1 & 0 & 0 \\ 0 & 1 & 0 \\ 0 & 0 & 0
\end{array}
\right),
\end{equation}
\begin{equation}
A^v=\left(
\begin{array}{ccc}
0 & 0 & 0 \\ 0 & 0 & 0 \\ 0 & 0 & Q
\end{array}
\right),
\end{equation}
and
\begin{equation}
A^\alpha=\left(
\begin{array}{ccc}
C^\alpha & 0 & 0 \\ 0 & C^\alpha & -\xi^\alpha \\ 0 & -\xib^\alpha & 0
\end{array}
\right).
\end{equation}
Whence the reduced system (\ref{M1})-(\ref{M3}) is clearly symmetric
hyperbolic \cite{Ste91}.

\begin{figure}[tbt]
\centering \mbox{ \epsfig{file=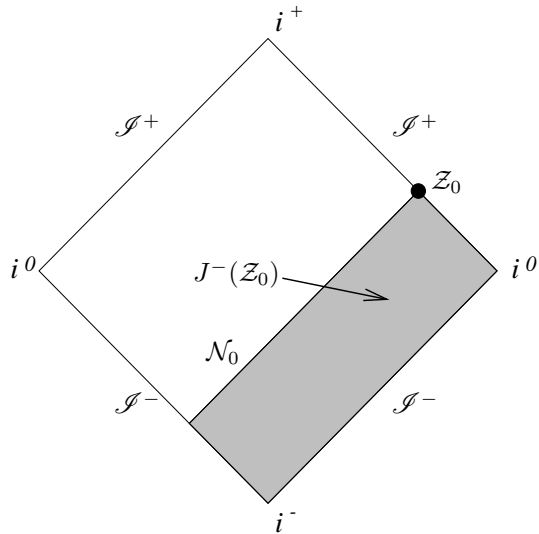,width=7cm,height=7cm}}
\put(-55,150){$\scri^+$}
\put(-55,45){$\scri^-$}
\put(-160,150){$\scri^+$}
\put(-160,45){$\scri^-$}
\put(-125,65){$\caln_0$}
\put(-40,130){$\calz_0$}
\put(-130,95){$J^-(\calz_0)$}
\caption{The asymptotic characteristic initial value problem for a zero rest-mass-field in an asymptotically simple spacetime.}
\end{figure}

Given suitable initial data on a future oriented null hypersurface
$\caln_0$ and on future null infinity $\scri^+$ one would like to
obtain to obtain the spin-1 zero-rest-mass field to the past of the
two initial hypersurfaces $J^-(\scri^+\cup\caln_0)$. This is the
\emph{asymptotic characteristic initial value problem}. Alternatively
one could give initial data on a past oriented null hypersurface
$\caln_0'$ and on $\scri^-$ and reconstruct the field on
$J^+(\scri^-\cup\caln_0')$. For the sake of concreteness, here I will
restrict to the first possibility.

From the structure of the principal part of the spin-1 zero-rest-mass
field equations, one finds that the right way of setting such an
\emph{asymptotic characteristic initial value problem}  for
$\phi_{AB}$ is set by prescribing:
\begin{enumerate}
\item $\phi_0$ on $\caln_0$,
\item $\phi_1$ on $\calz_0=\caln_0\cap\scri^+$,
\item and $\phi_2$ on $\scri^+$.
\end{enumerate}
If the initial data on $\caln_0$ ($\phi_0$) peels,
i.e. $\phit_0=\O(\Omega^3)$ or alternatively $\phi_0=\O(1)$), then
$\phi_0$ is bounded at $\scri^+$, and one has a \emph{regular}
asymptotic characteristic initial value problem. Furthermore, if the
data on $\caln_0$ is analytic with respect to v, and the data on
$\scri^+$ is analytic with respect to u, one can make use of Duff
\cite{Duf58} and Friedrich's techniques \cite{Fri82} \footnote{Duff
showed that the ideas of the Cauchy-Kovalevskaya theorem
\cite{CouHil62} for a Cauchy problem can be extended to the case ofa
characteristic initial value problem. In particular he considered
linear systems with data specified on an initial hypersurface which is
part characteristic and part spacelike. The construction for the full
chacteristic case has been done by Friedrich.}  theorem and prove that
there is a unique analytic ($C^\omega$) solution to the posed initial
value problem. The idea behind the proof is to construct another
system of partial differential equations whose solutions can be shown
to exist, be analytic and with an expansion that majorises the formal
series solutions of the original system of equations. This implies the
analyticity (and existence) of the solutions. In order to implement
this construction the different terms in the equations are required to
be analytic as well. The Duff-Friedrich's construction works for
quasilinear equations. Here, the situation is much simpler as the
zero-rest-mass field equations are linear.

 Further refinements can be implemented in order to show
existence for the case when the initial data is $C^\infty$ rather than
analytic \cite{Ren90,Kan96b}. In such a case the field is found to be
$C^\infty$. However, these further results may not concern us here.

\subsection{The non-regular initial value problem}

So much for the so-called regular initial value problem. What happens
if the initial data over $\caln_0$ is non-peeling, like for instance
$\phi_0=\phi_0^{2,0}v^{-1}+\cdots$? As discussed elsewhere
\cite{Val00a}, such initial data gives rise to logarithms in the
asymptotic expansions of the remaining components of the
zero-rest-mass field spinor $\phi_{AB}$, and in $\phi_0$ itself for
later times, even in the case where the logarithmic terms were not
present in the initial data. An initial value problem with this kind
of data that is not bounded at $\scri^+$, and henceforth it will be
called \emph{non-regular}.

In the case that concerns us here, that of a minimally polyhomogeneous
spin-1 zero-rest-mass field, the non-regular nature of $\phi_0$ and
$\phi_1$ at $\scri^+$ precludes us from setting an initial value
problem as the one discussed above, and using directly Duff and
Friedrich's techniques.

Assume for a moment that we have a polyhomogeneous spin-1
zero-rest-mass field. Due to the absolute convergence of the formal
series expansion, one can reshuffle the terms and write:
\begin{eqnarray}
&&\phi_0= \phii_0\ln v +\phio_0, \\ 
&&\phi_1= \phii_1\ln v +\phio_1, \\ 
&&\phi_2= \phii_2\ln v +\phio_2,
\end{eqnarray}
where the coefficients $\phii_n$ and $\phio_n$ are completely
logarithm free! Asymptotically one has that
\begin{equation}
\begin{array}{ccc}
\left\{ \begin{array}{l} \phiit_0=\O(\Omega^3) \\
 \phiit_1=\O(\Omega^2) \\ \phiit_2=\O(\Omega^2)  \end{array} \right.
 & \mbox{ or } & \left\{ \begin{array}{l} \phio_0=\O(1) \\
 \phio_1=\O(1) \\ \phio_2=\O(\Omega) \end{array} \right.
\end{array}
\end{equation}
and that
\begin{equation}
\begin{array}{ccc}
\left\{ \begin{array}{l} \phiot_0=\O(\Omega^2) \\
 \phiot_1=\O(\Omega^2) \\ \phiot_2=\O(\Omega^1)  \end{array} \right.
 & \mbox{ or } & \left\{ \begin{array}{l} \phio_0=\O(\Omega^{-1}) \\
 \phio_1=\O(1) \\ \phio_2=\O(1) \end{array} \right.
\end{array}
\end{equation}
for the other auxiliary field.

Thus, upon substitution on the field equations (\ref{M1})-(\ref{M3}),
one obtains the following equations for the \emph{auxiliary field},
\begin{eqnarray}
&&\Delta\!\!\phii_0-\delta\!\!\phii_1=(2\gamma-\mu)\phii_0-2\tau\phii_1+\sigma\phii_2,
\label{M11}\\
&&\Delta\!\!\phii_1-\delta\!\!\phii_2=-2\mu\phii_1+(2\beta-\tau)\phii_2, \label{M21}\\
&&D\!\!\phii_2-\deltab\!\!\phii_1=-\lambda\phii_0+2\pi\phii_1+\rho\phii_2, \label{M31}\\
&&D\!\!\phii_1-\deltab\!\!\phii_0=(\pi-2\alpha)\phii_0+2\rho\phii_1. \label{M41}
\end{eqnarray}
which are in fact identical to the spin-1 zero-rest-mass field
equations (\ref{M1})-(\ref{M4}). Again, the system is overdetermined, and thus one discards the last equation.

In an ordinary asymptotic characteristic initial value problem, the
initial data over $\caln_0$ is prescribed, that is
\begin{equation}
\phi_0=\phi_0^{4,0}v^{-1}+\left(\phi_0^{5,1}\ln
v+\phi_0^{5,0}\right)+\cdots,
\end{equation}
is given. Thus, the value of
\begin{equation}
\phii_0=\phi_0^{5,1}+\cdots,
\end{equation}
at $\caln_0$ is also known. Furthermore, as seen before
$\phi_1^{2,1}=\ethbar\phi_0^{2,0}$, and therefore we also know the
value of $\phii_1$ at $\calz_0$. Finally, as discussed
previously,$\phii_2=\O(v)$, and therefore $\phii_2=0$ on $\scri^+$,
i.e. the field is non-radiative \footnote{This means essentially that
the radiation field component $\phi_0$ has no type N term. Thus, it
does not contribute to the Poynting vector and consequently no energy
loss can be ascribed to it! }. Consequently, we are are in the
possession of all the ingredients needed to set an asymptotic
characteristic value problem for the spin-1 zero-rest-mass field
$\phii_{AB}$. The initial data has been constructed out of the
(non-regular) initial data for the original spin-1 zero-rest-mass
field $\phi_{AB}$. Duff and Friedrich's techniques allows us to state
that there is an (unique) analytic solution for this initial value
problem in a neighbourhood of $\calz_0$.

What happens with the other auxiliary field? In this case the
equations are,
\begin{eqnarray}
&&\Delta\!\!\phio_0-\delta\!\!\phio_1=(2\gamma-\mu)\phio_0-2\tau\phio_1+\sigma\phio_2,
\label{M10}\\
&&\Delta\!\!\phio_1-\delta\!\!\phio_2=-2\mu\phio_1+(2\beta-\tau)\phio_2,
\label{M20}\\
&&D\!\!\phio_2-\deltab\!\!\phio_1+v^{-1}Q\phii_2=-\lambda\phio_0+2\pi\phio_1+\rho\phio_2,
\label{M30} \\ &&D\!\!\phio_1-\deltab\!\!\phio_0+v^{-1}Q\phii_1
=(\pi-2\alpha)\phio_0+2\rho\phio_1. \label{M40}
\end{eqnarray}
Note the presence of the extra terms $v^{-1}Q\!\!\phii_2$ and
$v^{-1}Q\!\!\phii_1$. Thus the auxiliary field $\phio_{AB}$ satisfies
a kind of sourced Maxwell equations. The source happens to be
precisely the vacuum field $\phii_{AB}$. As it has been done before,
one discards the last equation (\ref{M40}).

From a previous analysis, one knows that $\phii_2=\O(v)$, and thus the
whole new term in equation (\ref{M30}) is such that
$v^{-1}Q\phii_2=\O(1)$. Hence, it is regular at $\scri^+$. However,
Duff and Friedrich's techniques cannot be applied to the reduced
system (\ref{M10})-(\ref{M30}), as it is, for the initial data over
the initial hypersurface $\caln_0$ is not regular.

As discussed, $\phio_0=\phi_0^{2,0}v^{-1}+\O(1)$. Substitution of this
into equation (\ref{M10}) yields,
\begin{equation}
\partial_u \phi_0^{2,0}=0.
\end{equation}
Therefore, the non-regular term is a constant of motion
\cite{Val00a}. Let $\phio_0=\phi_0^{2,0}v^{-1}+\phio_{0*}$, with
$\phio_{0*}=\O(1)$, i.e. from a physical spacetime point of view the
component $\phio_{0*}=\O(1)$ satisfies the peeling
behaviour. Substitution into equations (\ref{M1})-(\ref{M3}) yields,
\begin{eqnarray}
&&\Delta\!\!\phio_{0*}-\delta\!\!\phio_1=(2\gamma-\mu)\phi^{2,0}_0v^{-1}+(2\gamma-\mu)\phio_{0*}-2\tau\phio_1+\sigma\phio_2,
\label{M10star}\\
&&\Delta\!\!\phio_1-\delta\!\!\phio_2=-2\mu\phio_1+(2\beta-\tau)\phio_2,
\label{M20star}\\
&&D\!\!\phio_2-\deltab\!\!\phio_1+v^{-1}Q\phii_2=-\lambda\phi^{2,0}_0v^{-1}-\lambda\phio_{0*}+2\pi\phio_1+\rho\phio_2. \label{M30star}
\\
\end{eqnarray}
Now, $\gamma=\O(v)$, $\mu=\O(v)$, and $\lambda=\O(v)$. Therefore, the
new extra terms $(2\gamma-\mu)\phi^{2,0}_0v^{-1}$ and
$-\lambda\phi^{2,0}_0v^{-1}$ are regular at $\scri^+$ and
known. Moreover, they are also analytic at $v=0$. Therefore we have
obtained a perfectly regular symmetric hyperbolic system for
$\phio_{0*}$, $\phio_{1}$, and $\phio_2$.

Furthermore, we also know the value of $\phio_{0*}$ at the initial
hypersurface $\caln_0$ (obtainable from the non-regular initial value
problem for $\phi_{AB}$). The value of $\phio_1$ at $\calz_0$ is also
known. Finally $\phio_2\approx\phi_2$, and thus we also know the data
on $\scri^+$. Consequently we have obtained a regular system of
symmetric hyperbolic equations for the scalars $\phio_{0*}$,
$\phio_1$, $\phio_2$ with suitable initial data on which Duff and
Friedrich's ideas can be used. The conclusion is that there exists a
neighbourhood of $\calz_0$ on which there exists an (unique) analytic
solution ($C^\omega$) to the posed initial value problem. If
$\phio_{0*}$ is analytic in a neighbourhood of $\scri^+$ then clearly
$\phi_0=\phi_0^{2,0}v^{-1}$ has a convergent Laurent expansion on the
same neighbourhood.

So, we have seen that there exists a neighbourhood $\scru_1$ of
$\calz_0$ for which the field $\phii_{AB}$ is analytic (in $v$), and
another neighbourhood $\scru_0$ for which the field $\phio_{AB}$ has a
Laurent expansion (again in $v$). Let
$\scru=\scru_0\cap\scru_1$. Thus, one can say that the non-regular
asymptotic characteristic initial value problem for the zero-rest-mass
field $\phi_{AB}=\phio_{AB}+\phii_{AB}\ln v$ has a (unique) solution
in $\scru\!\!-\!\!\scri^+$ which is in fact polyhomogeneous. Note that
althought the field is not well defined at $\scri^+$, the radiation
field is well defined ($\phi_2$) as it is part of the regular initial
data. Furthermore, it can be shown that the net electromagnetic charge
of the field (as measured from null infinity) is also well defined.

These results can be sumarised in the form of the following

\begin{theorem}
Let $(\scrmtilde,\gtilde_{ab})$ be a Ricci flat ($R_{ab}=0$) asymptotically simple spacetime, and let $(\scrm,g_{ab})$ be the corresponding unphysical spacetime obtained by the conformal rescaling. Furthermore, let $\caln_0$ be a future oriented null hypersurface in $\scrm$ that intersects $\scri^+$ at $\calz_0$. Then there exist a neighbourhood $\scru$ of $\calz_0$ such that the non-regular asymptotic characteristic initial value problem for a spin-1 zero-rest-mass field with minimally polyhomogeneous data on $\caln_0$ has a unique (polyhomogeneous) solution in $\scru\!\!-\!\!\scri^+$.  
\end{theorem}

\section{Conclusions}
An existence result for the asymptotic characteristic initial value
problem for a spin-1 zero-rest-mass field with polyhomogeneous initial
data has been proved. The polyhomogeneous initial data is not bounded
at $\scri^+$, and thus, the initial value problem is non-regular. This
means essentially that standard existence theorems for symmetric
hyperbolic equations cannot be directly used. However, as seen it
possible to reformulate this non-regular initial value problem as two
coupled initial value problems for some auxiliary fields. These
auxiliary fields satisfy partial differential equations with regular
initial data. Moreover, these auxiliary systems have the same
principal part (symbol) as the spin-1 zero-rest-mass field. 

The existence result shows that the polyhomogeneous structure of the
initial data is preserved at least in a neighbourhood of
$\calz_0$. And for the same price, one also settles the matter of the
convergence of polyhomogeneous expansions for zero-rest-mass fields.

\subsection{Further extensions}
The proof given here is limited to minimally polyhomogeneous spin-1
zero-rest-mass field equations. A proof for an arbitrary
polyhomogeneous spin-2 zero-rest-mass field (higher spins are not so
interesting as the Buchdahl constraint has to be satisfied) follows
using the same techniques. Arbitrary polyhomogeneity in the initial
data means that further auxiliary fields and their respective systems
should be included in the discussion. As it has been pointed out
elsewhere, these auxiliary fields form a hierarchy. This hierarchy is
such that the field equations for a given auxiliary field has the
auxiliary field just above in the hierarchy as source term (cfr. with
the equations for $\phii_{AB}$ and $\phio_{AB}$). The consideration of
a spin-2 field rather than a spin-1 one would only increase the amount
of algebra involved.

A much more interesting and natural extension of this work is to
trying to apply similar techniques to polyhomogeneous gravitational
fields. This would require the use of Friedrich's regular conformal
field equations, and a good understanding of the formal
polyhomogeneous solutions to the aforementioned equations. These
concerns will be and are the concern of future work.
  
\section*{Acknowledgements}
I want to thank my supervisor Prof. M.A.H. MacCallum for his advice
and encouragement, and for suggesting me to work on the topics
discussed here and in previous articles. I hold a scholarship
(110441/110491), from the Consejo Nacional de Ciencia y
Tecnolog\'{\i}a (CONACYT), Mexico.

\end{document}